\documentclass{article}

\usepackage{PRIMEarxiv}
\usepackage{graphicx}%
\usepackage{multirow}%
\usepackage{amsmath,amssymb,amsfonts}%
\usepackage{amsthm}%
\usepackage{mathrsfs}%
\usepackage[title]{appendix}%
\usepackage{xcolor}%
\usepackage{textcomp}%
\usepackage{manyfoot}%
\usepackage{booktabs}%
\usepackage{algorithm}%
\usepackage{algorithmicx}%
\usepackage{algpseudocode}%
\usepackage{listings}%
\usepackage{verbatim}

\PassOptionsToPackage{numbers,sort&compress}{natbib}
\usepackage{natbib}

\usepackage[utf8]{inputenc} 
\usepackage[T1]{fontenc}    
\usepackage{hyperref}       
\usepackage{url}            
\usepackage{booktabs}       
\usepackage{amsfonts}       
\usepackage{nicefrac}       
\usepackage{microtype}      
\usepackage{lipsum}
\usepackage{fancyhdr}       
\usepackage{graphicx}       
\graphicspath{{media/}}     

\title{Scalable Dielectric Tensor Predictions for Inorganic Materials using Equivariant Graph Neural Networks}

\author{
  Haowei Hua$^{1}$, Chen Liang$^{2}$, Ding Pan$^{3,4}$, Irwin King$^{5}$, Shengchao Liu$^{5}$, Koji Tsuda$^{2,6,7}$, Wanyu Lin$^{1,8}$\thanks{Corresponding author: \texttt{wan-yu.lin@polyu.edu.hk}} 
  \\
  \normalfont\small
  $^{1}$Department of Computing, The Hong Kong Polytechnic University,
  Hong Kong, China \\
  \normalfont\small
  $^{2}$Department of Computational Biology and Medical Sciences, 
  Graduate School of Frontier Sciences, \\
  \normalfont\small
  The University of Tokyo,
  5-1-5 Kashiwanoha, Kashiwa 277-8561, Japan \\
  \normalfont\small
  $^{3}$Department of Physics, Hong Kong University of Science and Technology,
  Hong Kong, China \\
  \normalfont\small
  $^{4}$Department of Chemistry, Hong Kong University of Science and Technology,
  Hong Kong, China \\
  \normalfont\small
  $^{5}$Department of Computer Science and Engineering, 
  The Chinese University of Hong Kong,
  Hong Kong, China \\
  \normalfont\small
  $^{6}$Center for Basic Research on Materials, National Institute for Materials Science,
  1-1 Namiki, Tsukuba, Ibaraki 305-0044, Japan \\
  \normalfont\small
  $^{7}$RIKEN Center for Advanced Intelligence Project,
  1-4-1 Nihonbashi, Chuo-ku, Tokyo 103-0027, Japan \\
  \normalfont\small
  $^{8}$Department of Data Science and Artificial Intelligence,
  The Hong Kong Polytechnic University,
  Hong Kong, China
}

\begin{document}
\maketitle

\begin{abstract}
Accurate prediction of dielectric tensors is essential for accelerating the discovery of next-generation inorganic dielectric materials. Existing machine learning approaches, such as equivariant graph neural networks, typically rely on specially-designed network architectures to enforce O(3) equivariance. However, to preserve equivariance, these specially-designed models restrict the update of equivariant features during message passing to linear transformations or gated equivariant nonlinearities. The inability to implicitly characterize more complex nonlinear structures may reduce the predictive accuracy of the model. In this study, we introduce a frame-averaging-based approach to achieve equivariant dielectric tensor prediction. We propose GoeCTP, an O(3)-equivariant framework that predicts dielectric tensors without imposing any structural restrictions on the backbone network. We benchmark its performance against several state-of-the-art models and further employ it for large-scale virtual screening of thermodynamically stable materials from the Materials Project database. GoeCTP successfully identifies various promising candidates, such as Zr(InBr$_3$)$_2$ (band gap $E_g = 2.41$ eV, dielectric constant $\overline{\varepsilon} = 194.72$) and SeI$_2$ (anisotropy ratio $\alpha_r = 96.763$), demonstrating its accuracy and efficiency in accelerating the discovery of advanced inorganic dielectric materials.
\end{abstract}





\maketitle

\section{Introduction}\label{sec1}

The dielectric tensor $\boldsymbol{\varepsilon} \in \mathbb{R}^{3 \times 3}$ is a second-order physical quantity that characterizes a material’s polarization response under an external electric field. 
Materials with dielectric tensors satisfying specific requirements serve as essential building blocks in numerous modern technologies, including computer memory \cite{hwang2015prospective,yim2015novel}, energy storage \cite{wu2022advanced,li2018high}, and microwave communication \cite{hill2021perspective,shehbaz2023recent}.
Therefore, accurately predicting the dielectric tensor is critical, as it supports the rational design and accelerated discovery of high-performance dielectric materials. Traditionally, such predictions heavily rely on first-principles simulations, particularly those based on density functional theory (DFT) \cite{hautier2012computer}. These approaches require solving the computationally expensive Kohn-Sham equations \cite{hohenberg1964inhomogeneous, kohn1965self}. The associated computational cost becomes prohibitive for high-throughput screening across large-scale materials databases, forming a significant bottleneck in the discovery of novel dielectric materials.

Recent advancements in machine learning (ML) offer promising alternatives \cite{xie2018crystal,louis2020graph,choudhary2021atomistic,kaba2022equivariant,yan2022periodic,lin2023efficient,das2023crysgnn,lee2023density,kang2023multi,gong2023examining,vu2023towards,banik2023cegann,yancomplete,taniaicrystalformer,wang2024conformal,song2024diffusion,gupta2024structure,chen2024learning,koker2024higher,li2024md,okabe2024virtual,zou2025predicting,ito2025rethinking,jin2025transformer,jasperson2025cross,shen2025denoising,chen2025multi,madani2025accelerating,noda2025advancing,li2025probing,wang2025machine,gibson2025accelerating,wu2025hierarchy,huangcode,song2025accurate,niyongabo2025llm,hua2025local}, enabling the approximation of DFT-level accuracy with substantially reduced computational cost. Specifically, ML-based approaches directly learn a mapping from material structures to dielectric tensor properties, thereby bypassing the need for solving the Kohn-Sham equations. With ML, predictions could be obtained through efficient neural network inference, primarily involving matrix operations.
Among existing ML approaches, equivariant graph neural networks (EGNNs) have shown strong potential for predicting tensorial material properties \cite{zhong2023general,yan2024space,mao2024dielectric,wen2024equivariant}. These models are carefully designed to respect the physical symmetries of atomic systems, particularly O(3) equivariance. For instance, ETGNN \cite{zhong2023general} represents the tensor as a linear combination of local spatial features projected along edge directions, ensuring O(3) equivariance. GMTnet \cite{yan2024space} incorporates directional features and spherical harmonics, employing tensor products to preserve O(3) equivariance while reducing symmetry constraints to a crystal-level representation.
More recently, DTNet \cite{mao2024dielectric} has emerged as a representative model for dielectric tensor prediction. It constructs rank-2 equivariant outputs by combining tensor features with the outer product of vector features. In addition, DTNet employs the PreFerred Potential (PFP) \cite{takamoto2022towards} model as a pretrained backbone. PFP was trained on 22 million DFT-calculated potential energies and serves as a multi-order encoder capable of capturing universal compositional and structural information. By exploiting this rich latent representation, DTNet achieves superior predictive performance, even under data-scarce conditions.

Despite these advances, EGNN approaches such as DTNet still exhibit notable limitations.
Specifically, to maintain equivariance, existing methods including DTNet and GMTNet rely on specially designed message-passing mechanisms. These architectural constraints limit scalability and hinder the ability to capture complex chemical bonds and many-body interactions that are prevalent in crystalline systems. For instance, DTNet generates a rank-2 equivariant dielectric tensor by combining the outer product of vector features with a second-order Euclidean tensor. To preserve equivariance, DTNet restricts the update of both vector and Euclidean tensor features during message passing to linear transformations or gated equivariant nonlinearities. The inability to incorporate more complex nonlinear structures may reduce the predictive accuracy of the model. Consequently, their effectiveness in large-scale materials screening remains limited. 
In addition, DTNet relies on the PFP module, which is a proprietary and closed source commercial model. This dependency restricts further methodological development.

To address these challenges, we propose GoeCTP, a scalable O(3)-equivariant framework designed for accurate dielectric tensor prediction. Our approach adopts frame-averaging strategies \cite{ma2024canonicalization,duval2023faenet,linequivariance,kaba2023equivariance,dymequivariant,punyframe} to enforce O(3) equivariance. 
Because this strategy imposes no restrictions on backbone network design, GoeCTP can flexibly incorporate scalable message-passing mechanisms, achieving high predictive accuracy.
Experimental results show that GoeCTP achieves state-of-the-art performance across multiple benchmark datasets. Furthermore, it has been successfully applied to a high-throughput screening task involving more than 10,000 candidate materials. The screening process identified several novel compounds exhibiting strong average dielectric responses or high dielectric anisotropies. These results demonstrate the practical potential of GoeCTP in accelerating the discovery of functional dielectric materials.


\begin{figure*}[t] 
  \centering   
  \includegraphics[width=\textwidth]{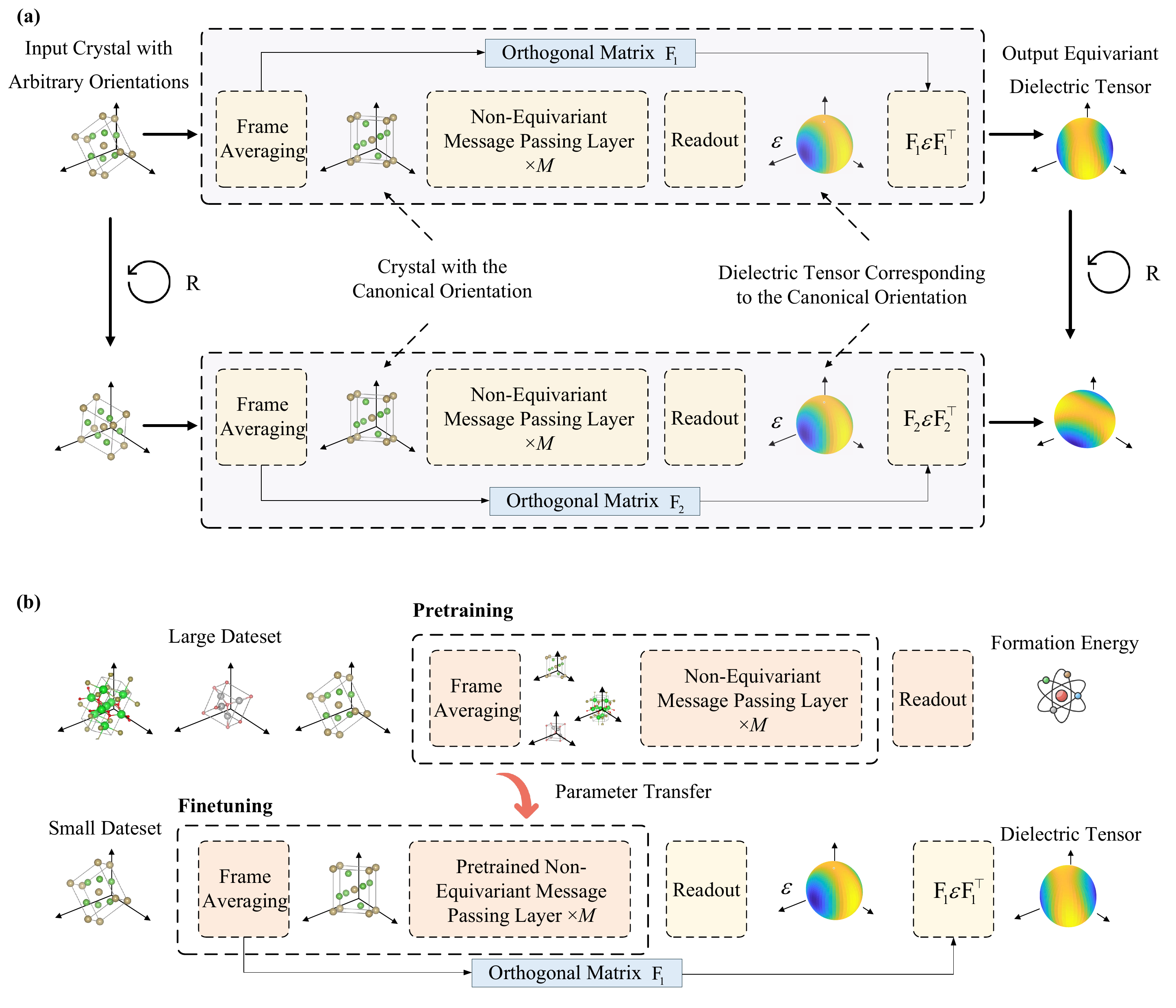}
  \caption{(a) Illustration of GoeCTP. Given an input crystal, GoeCTP first applies frame averaging to align the crystal structure to its canonical orientation. A crystal graph is then constructed, and the message passing layers together with the readout module generate a predicted dielectric tensor $\boldsymbol{\varepsilon}$ corresponding to this canonical orientation. The frame $\text{F}_1$ computed at the input stage is subsequently applied to this canonical tensor to obtain the final tensor prediction. When an arbitrary rotation or reflection $\text{R}$ is applied to the input crystal structure, frame averaging again aligns the transformed crystal to the same canonical orientation. In this case, the resulting frame satisfies $\text{F}_2=\text{R}\text{F}_1$. Consequently, at the output stage, the predicted tensor is transformed consistently, ensuring that the final output satisfies the required O(3) equivariance. (b) Diagram of the GoeCTP pretraining strategy.
During pretraining on a scalar property dataset, the frame obtained through frame averaging is not applied to the output. After pretraining is completed, the readout module is replaced, and the message passing layers are initialized with the pretrained weights. The model is then fine-tuned on the tensor dataset.}
  \label{fig:overview}
\end{figure*}

\section{Results}\label{sec2}

\subsection{Frame averaging}
In crystal tensor prediction tasks, the model output is typically required to satisfy O(3) equivariance.
Formally, let the atomic positions of a crystal structure be denoted as 
$\mathbf{X}$, and let $f_\theta(\cdot)$ represent the prediction model with $f_\theta(\mathbf{X})=\boldsymbol{\varepsilon}$. The equivariance requirement can be expressed as follows:
\begin{equation}
\label{equivariance_eq}
f_\theta(\mathbf{R}\mathbf{X}) = \mathbf{R}f_\theta(\mathbf{X})\mathbf{R}^\top,
\end{equation}
where $\mathbf{R}\in\mathbb{R}^{3 \times 3}$ is an arbitrary O(3) group transformation. 

To satisfy this requirement, we start from an arbitrary non-equivariant prediction model $g_\theta(\cdot)$ and construct an equivariant model $f_\theta(\cdot)$ through frame averaging \citep{linequivariance, ma2024canonicalization}.
Here, a frame can be interpreted as an orthogonal matrix $\mathbf{F}\in\mathbb{R}^{3 \times 3}$,
obtained from an O(3)-equivariant mapping $\mathbf{F}=h(\mathbf{X})$.
The equivariant model is then defined as:
\begin{equation}
f_\theta(\mathbf{X})=\mathbf{F}g_\theta(\mathbf{F}^\top\mathbf{X})\mathbf{F}^\top.
\label{frame_qian}
\end{equation}
For any rotation or reflection $\mathbf{R}$, since $\mathbf{F}$ is equivariant, we have
\begin{equation}
f_\theta(\mathbf{R}\mathbf{X})=\mathbf{R}\mathbf{F}g_\theta((\mathbf{R}\mathbf{F})^\top\mathbf{R}\mathbf{X})(\mathbf{R}\mathbf{F})^\top=\mathbf{R}\mathbf{F}g_{\theta}(\mathbf{F}^\top\mathbf{X})\mathbf{F}^\top\mathbf{R}^\top=\mathbf{R}f_\theta(\mathbf{X})\mathbf{R}^\top.
\label{gx_qian}
\end{equation}
This formulation completely decouples the requirement of equivariance from the architectural design of the neural network $g_\theta(\cdot)$, allowing message-passing mechanisms to be flexibly designed without any additional symmetry constraints.
In this work, we employ polar decomposition \citep{higham1986computing, jiaospace, hall2013lie} as the mapping $h(\cdot)$ to compute the frame $\mathbf{F}$. Further details are provided in the Method section.

\subsection{Overview of GoeCTP}
The overall architecture of GoeCTP is depicted in Fig.~\ref{fig:overview}.
GoeCTP enforces equivariance through frame averaging, which introduces no restrictions on the design of the message passing module. This flexibility allows the use of arbitrary message passing strategies to enhance predictive performance, in contrast to DTNet, which restricts the update of both vector and Euclidean tensor features during message passing to linear transformations or gated equivariant nonlinearities.

In this study, we present three non-equivariant message passing schemes, and the corresponding networks are referred to as GoeCTP, GoeCTP (Mat.), and GoeCTP (iCom.). 
The base GoeCTP model employs a single non-equivariant message passing layer as its backbone architecture. GoeCTP (Mat.) integrates a transformer-based message passing module proposed in Matformer \citep{yan2022periodic}. Building upon this design, GoeCTP (iCom.) further extends GoeCTP (Mat.) by incorporating an additional edge feature update layer introduced in iComFormer \citep{yancomplete}.
Following the multi-edge crystal graph construction used in CGCNN \cite{xie2018crystal}, let $v^{(l)}_i$ denote the atomic feature vector of the $i$-th atom at layer $l$, and let $e^{(h)}_{ij}$ represent the $h$-th edge feature between atoms $i$ and $j$. 
The message passing process in GoeCTP is defined as
\begin{equation}
\begin{aligned}
& \boldsymbol{msg}_{ij}=\sum_{h}\boldsymbol{\xi}(\operatorname{LN} (v^{(l)}_i)|\operatorname{LN} (v^{(l)}_j)|\operatorname{LN}_e (e^{(h)}_{ij})), \\
&v^{(l+1)}_i=\operatorname{softplus}(v^{(l)}_i+\mathrm{BN}(\sum_{j\in\mathcal{N}_i}\boldsymbol{msg}_{ij})), \\
\end{aligned}
\end{equation}
where $\mathrm{LN}(\cdot)$ and $\mathrm{LN}_e(\cdot)$ denote the linear transformation layers, $\boldsymbol{\xi}(\cdot)$ represents a nonlinear transformation function, $\operatorname{BN}(\cdot)$ refers to the batch normalization, and $|$ denote the concatenation. Further details on GoeCTP (Mat.) and GoeCTP (iCom.) are provided in the Method section.

Moreover, since DTNet relies on a pretrained PFP module to enhance its performance, we also examine the impact of pretraining and fine-tuning strategies on the accuracy of GoeCTP. Specifically, we modify the readout layer of GoeCTP to output a scalar and pretrain the model on a large scalar property dataset. After pretraining, we retain the pretrained parameters of the non-equivariant message passing layers as initialization, replace the readout layer with one that outputs a tensor, and then fine-tune the entire model on the dielectric tensor dataset by updating all parameters.

\subsection{JARVIS-DFT Dataset}
To comprehensively evaluate the performance of the proposed GoeCTP, we conduct comparative experiments against a variety of baseline models across different datasets. In this section, we present preliminary comparisons on a dielectric tensor dataset derived from the JARVIS-DFT database. The detailed experimental procedure is described as follows.

\textbf{Data preparation.}
We directly adopt the dielectric tensor dataset curated and constructed by \citet{yan2024space}, which is derived from one of the most widely used open databases, the JARVIS-DFT database \cite{choudhary2020joint}. This dataset extracts tensor property values and corresponding crystal structures directly from DFT calculation files. Such an extraction strategy ensures that the symmetry of the dielectric tensor remains consistent with the symmetry of the underlying crystal structure.
The dataset contains a total of 4911 crystal structures, each annotated with its corresponding dielectric tensor. It spans more than 80 distinct chemical elements and covers all seven crystal systems. The maximum absolute value of all scalar values within the dielectric tensors is constrained to be less than 1000.
For experimental evaluation, the dataset is divided into training, validation, and test sets with a ratio of 0.8 to 0.1 to 0.1, respectively. To mitigate the effects of randomness, each experiment is independently repeated five times with different random data splits. 
The mean absolute error (MAE) metric is employed to evaluate the accuracy of the predicted dielectric tensors. The final results are reported as the mean and standard deviation across these five runs.

\textbf{Model comparison for dielectric tensor prediction.}
In this work, we select several advanced methods as baseline models for comparison, including ETGNN \cite{zhong2023general}, GMTNet \cite{yan2024space}, and MatTen \cite{wen2024equivariant}. 
These methods employ specially designed message passing strategies for equivariant tensor prediction, such as the incorporation of tensor product operations. Therefore, they are suitable benchmarks for evaluating our GoeCTP framework, which allows flexible message passing design.
For our proposed GoeCTP framework, we evaluate three distinct non-equivariant message passing variants: GoeCTP, GoeCTP (Mat.), and GoeCTP (iCom.). Full architectural details are provided in the Methods section.

Table \ref{dielectric_JARVIS} presents a comprehensive comparison between the three baseline models and our proposed GoeCTP variants. For the cross-system prediction task, all models were trained on the complete training dataset, while test structures were partitioned according to their respective crystal systems for evaluation.
The experimental results show that both GoeCTP (Mat.) and the base GoeCTP model achieve the lowest prediction errors in two out of seven crystal systems, demonstrating their competitive advantage over the baseline models. This advantage highlights the effectiveness of flexible message-passing design strategies. On the overall test set, GoeCTP exhibits significantly lower error compared to all other methods, despite using only a single message-passing layer. GoeCTP (Mat.) also consistently outperforms all baseline algorithms. In contrast, GoeCTP (iCom.) exhibits relatively weaker performance compared to the other two variants.
The superior performance of the simpler message-passing strategies on the JARVIS-DFT dataset may be attributed to the limited dataset size. More complex architectures are likely to overfit smaller datasets, which can reduce generalization capability. Furthermore, all methods show considerable variance across five independent runs with different dataset partitions. This variability, together with the close mean performance among models, may influence the statistical confidence of the results. To further verify these findings and improve robustness, additional experiments are being conducted on the Materials Project dataset.

\begin{table*}[htbp]
\centering
\caption{Performance comparison of models across all crystal systems on the JARVIS-DFT dielectric dataset.}
\scalebox{0.82}{
\begin{tabular}{l c c c c c c}
\toprule
 \textbf{System}  & \textbf{GoeCTP} &  \textbf{GoeCTP (Mat.)} &\textbf{GoeCTP (iCom.)} & \textbf{GMTNet}& \textbf{ETGNN}  & \textbf{MatTen} \\
\midrule

Cubic        & 4.838 $\pm$ 1.926 & 4.649 $\pm$ 1.846 & 4.277 $\pm$ 0.900 & \textbf{4.158 $\pm$ 0.968} & 3.799 $\pm$ 0.830 & 4.172 $\pm$ 0.844 \\

Hexagonal    & 2.099 $\pm$ 1.245 & 2.253 $\pm$ 1.349 & 2.551 $\pm$ 1.596 & 2.207 $\pm$ 1.167 & \textbf{1.504 $\pm$ 0.440 }& 2.292 $\pm$ 1.056 \\

Monoclinic   & 2.376 $\pm$ 0.858 & 2.457 $\pm$ 0.984 & 2.574 $\pm$ 1.196 & 2.626 $\pm$ 0.949 & 2.917 $\pm$ 1.314 &\textbf{ 2.250 $\pm$ 0.927} \\

Orthorhombic & \textbf{2.112 $\pm$ 1.074} & 2.327 $\pm$ 1.159 & 2.775 $\pm$ 0.718 & 2.498 $\pm$ 1.210 & 2.168 $\pm$ 1.209 & 2.370 $\pm$ 1.074 \\

Tetragonal   & \textbf{2.971 $\pm$ 2.006} & 3.593 $\pm$ 2.109 & 3.203 $\pm$ 2.093 & 3.232 $\pm$ 2.068 & 3.910 $\pm$ 1.301 & 3.138 $\pm$ 1.944 \\

Triclinic    & 3.917 $\pm$ 2.194 & \textbf{3.785 $\pm$ 2.123 }& 4.516 $\pm$ 1.582 & 4.232 $\pm$ 2.233 & 5.089 $\pm$ 1.933 & 4.654 $\pm$ 2.219 \\

Trigonal     & 3.747 $\pm$ 1.730 & \textbf{3.703 $\pm$ 1.288} & 4.147 $\pm$ 1.568 & 4.483 $\pm$ 2.152 & 3.992 $\pm$ 2.232 & 4.410 $\pm$ 1.385 \\

All          &\textbf{ 3.197 $\pm$ 0.689} & 3.290 $\pm$ 0.794 & 3.441 $\pm$ 0.556 & 3.410 $\pm$ 0.766 & 3.343 $\pm$ 0.548 & 3.337 $\pm$ 0.510 \\
\bottomrule
\end{tabular}
}
\footnotetext{Source: This is an example of table footnote. This is an example of table footnote.}
\label{dielectric_JARVIS}
\end{table*}

\subsection{Materials Project Dataset}
\label{Materials_Project_Dataset}

\textbf{Data preparation.} 
In addition to the preliminary experiments conducted on the dielectric tensor dataset from the JARVIS-DFT database, we further perform a more comprehensive evaluation of GoeCTP on a dielectric tensor dataset derived from the Materials Project database. In this section, we also include the recently proposed DTNet \citet{mao2024dielectric} as an additional baseline for comparison.
Due to the proprietary and closed-source nature of the PFP module used in DTNet, we are unable to reproduce its implementation. To ensure a fair comparison, we employ the dielectric tensor dataset curated by \citet{mao2024dielectric}, which is constructed from one of the most widely used open-access databases, the Materials Project database \cite{jain2013commentary}, and was also used in the DTNet study. 
This dataset contains 6648 crystal structures along with their corresponding dielectric tensor labels. It includes 72 distinct chemical elements and spans all seven crystal systems. The scalar values in the dielectric tensors range from 1.155 to 98.889.
Following the same experimental protocol as used for the JARVIS-DFT dataset, the Materials Project dataset is divided into training, validation, and testing sets in a ratio of 0.8 to 0.1 to 0.1. To mitigate the effects of randomness, each experiment is independently repeated five times with different random data splits. The final results are reported as the mean and standard deviation across these five runs. It is worth noting that the experimental setup in this study is consistent with the configuration reported in the DTNet paper.

\textbf{Model comparison for dielectric tensor prediction.}
The GoeCTP variants consistently demonstrate superior performance compared to the baseline models, further validating the effectiveness of flexible message-passing design strategies.
Within the GoeCTP family, both GoeCTP and GoeCTP (iCom.) achieve the lowest prediction errors in three out of seven crystal systems while maintaining comparable performance on the overall test set. These results confirm the robustness and adaptability of the proposed framework across diverse crystal systems.
To enable more comprehensive and reliable evaluation, subsequent experimental analyses will be conducted using the Materials Project dataset.

\begin{table*}[htbp]
\centering
\caption{Performance comparison of models across all crystal systems on the Materials Project dielectric dataset.}
\scalebox{0.78}{
\begin{tabular}{l c c c c c  c c}
\toprule
 \textbf{System}  & \textbf{GoeCTP} &  \textbf{GoeCTP (Mat.)} &\textbf{GoeCTP (iCom.)}& \textbf{DTNet}  & \textbf{GMTNet}& \textbf{ETGNN}  & \textbf{MatTen} \\
\midrule
Cubic        & 1.444 $\pm$ 0.140 & 1.469 $\pm$ 0.213 &\textbf{ 1.363 $\pm$ 0.130} & 1.808 $\pm$ 0.615 & 1.369 $\pm$ 0.232 & 1.537 $\pm$ 0.215 & 2.548 $\pm$ 0.845 \\
Hexagonal    & 1.307 $\pm$ 0.300 & 1.314 $\pm$ 0.323 & \textbf{1.187 $\pm$ 0.343} & 1.409 $\pm$ 0.199 & 1.463 $\pm$ 0.505 & 1.547 $\pm$ 0.459 & 1.954 $\pm$ 0.266 \\
Monoclinic   & \textbf{1.524 $\pm$ 0.188 }& 1.570 $\pm$ 0.192 & 1.620 $\pm$ 0.190 & 1.854 $\pm$ 0.098 & 1.802 $\pm$ 0.213 & 1.656 $\pm$ 0.188 & 2.267 $\pm$ 0.184 \\
Orthorhombic & \textbf{1.394 $\pm$ 0.082} & 1.418 $\pm$ 0.095 & 1.404 $\pm$ 0.147 & 2.141 $\pm$ 0.089 & 1.620 $\pm$ 0.125 & 1.546 $\pm$ 0.127 & 2.461 $\pm$ 0.180 \\
Tetragonal   & 1.428 $\pm$ 0.240 &\textbf{ 1.405 $\pm$ 0.301 }& 1.417 $\pm$ 0.312 & 2.018 $\pm$ 0.373 & 1.457 $\pm$ 0.279 & 1.558 $\pm$ 0.291 & 2.351 $\pm$ 0.434 \\
Triclinic    & 1.822 $\pm$ 0.290 & 1.918 $\pm$ 0.266 & \textbf{1.799 $\pm$ 0.344 }& 2.142 $\pm$ 0.105 & 2.293 $\pm$ 0.237 & 1.999 $\pm$ 0.284 & 2.488 $\pm$ 0.136 \\
Trigonal     & \textbf{1.128 $\pm$ 0.091} & 1.187 $\pm$ 0.084 & 1.148 $\pm$ 0.089 & 1.739 $\pm$ 0.269 & 1.254 $\pm$ 0.242 & 1.200 $\pm$ 0.065 & 2.040 $\pm$ 0.371 \\
All          & \textbf{1.411 $\pm$ 0.064} & 1.441 $\pm$ 0.052 & 1.414 $\pm$ 0.067 & 1.911 $\pm$ 0.094 & 1.567 $\pm$ 0.031 & 1.541 $\pm$ 0.063 & 2.320 $\pm$ 0.153 \\
\bottomrule
\end{tabular}
}
\footnotetext{Source: This is an example of table footnote. This is an example of table footnote.}
\label{dielectric_MP}
\end{table*}

\begin{figure*}[t] 
  \centering   
  \includegraphics[width=\textwidth]{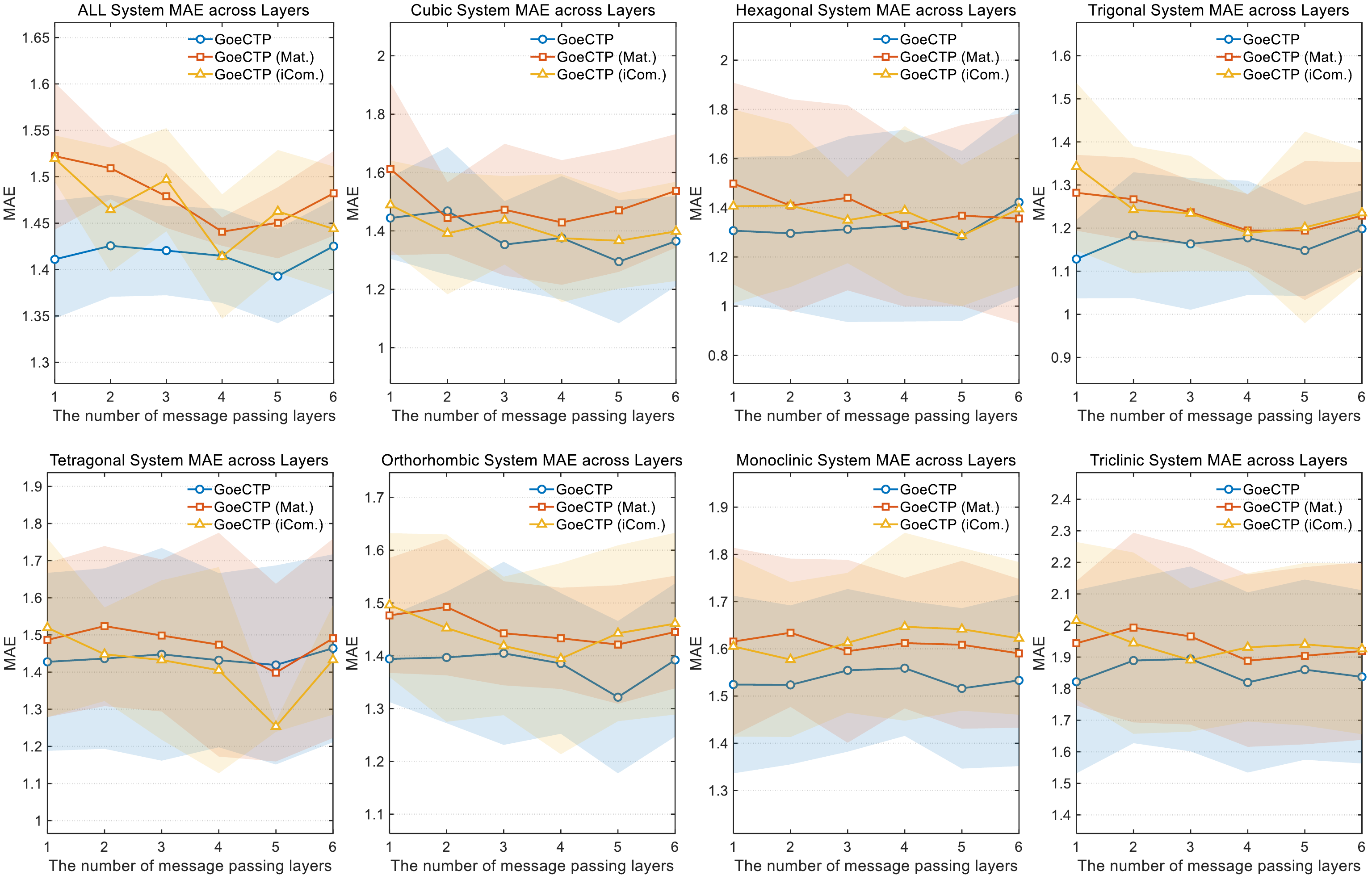}
  \caption{Impact of message passing depth on network performance. For each crystal system, the mean absolute error is reported (points), and the shaded regions denote standard deviations across multiple runs, enabling comparison among GoeCTP, GoeCTP (Mat.), and GoeCTP (iCom.).
  } 
  \label{different_layer}
\end{figure*}

\textbf{Performance of GoeCTP.}
In addition to comparisons with other baseline models, we conducted a series of detailed experiments focusing specifically on GoeCTP. In the previous comparison among GoeCTP, GoeCTP (Mat.), and GoeCTP (iCom.), the single-layer GoeCTP demonstrated the best predictive performance. Motivated by this observation, we further investigated how varying the number of message-passing layers influences the network’s performance. The corresponding experimental results are presented in Figure \ref{different_layer}.
Based on the mean MAE across the entire test set, the performance of GoeCTP remains largely stable with respect to the number of message-passing layers, except for the five-layer configuration, which slightly outperforms the single-layer version. In contrast, the performance of GoeCTP (Mat.) and GoeCTP (iCom.) exhibits stronger sensitivity to the number of layers, with both models achieving their best results when using four message-passing layers.
The optimal configuration may vary depending on the specific crystal system under consideration.

\begin{figure*}[t] 
  \centering   
  \includegraphics[width=\textwidth]{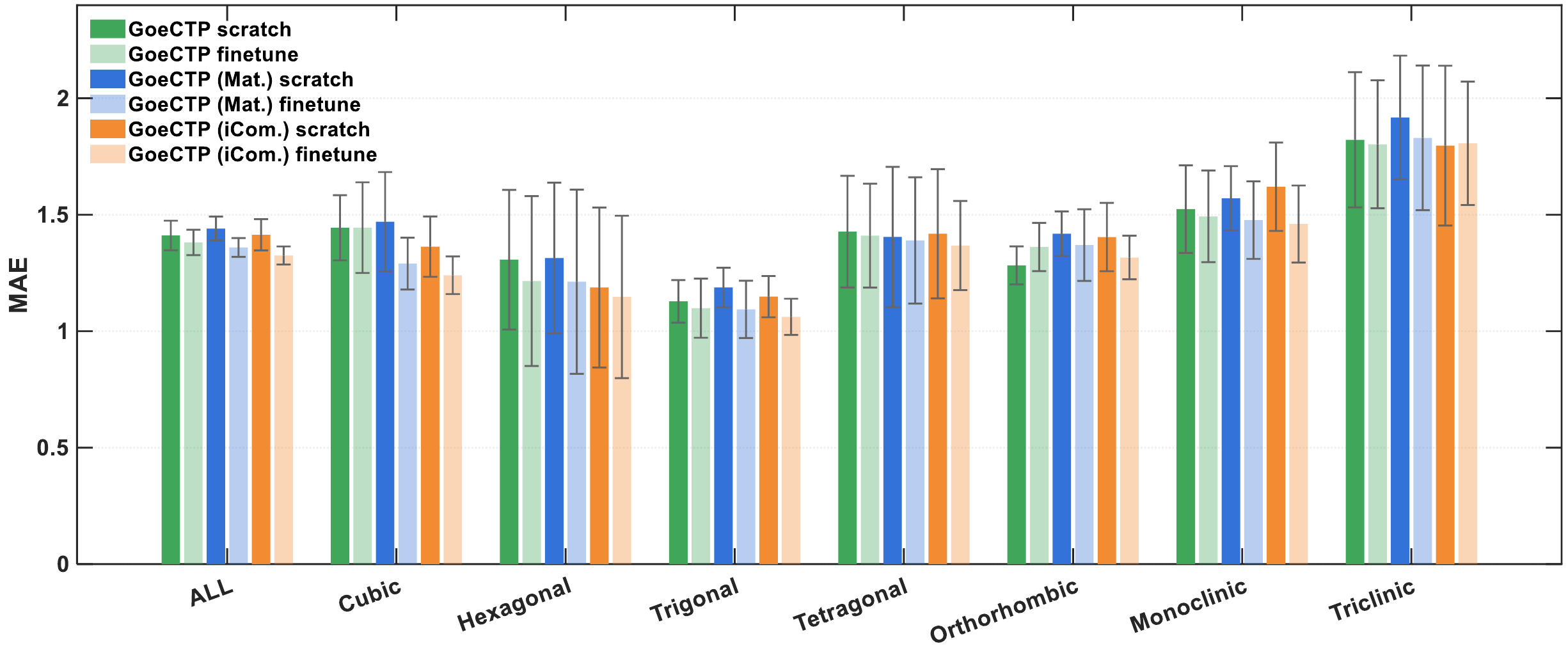}
  \caption{Comparison of model performance between training from scratch and fine tuning. Training from scratch indicates that the models are trained solely on the tensor datasets, while fine tuning corresponds to first pretraining GoeCTP on a large scalar dataset followed by fine tuning on the tensor datasets.
  } 
  \label{dielectric_MP_finetuning}
\end{figure*}

As noted in the DTNet study \cite{mao2024dielectric}, a fundamental challenge in applying EGNNs to materials science lies in their substantial data requirements. The availability of labeled data varies considerably across different material properties. For instance, while the Materials Project database contains more than 150,000 DFT-relaxed crystal structures with formation energy annotations, only about 7,000 entries (no more than 5\%) provide corresponding dielectric tensor data \cite{jain2013commentary,mao2024dielectric}.
To address the issue of data scarcity, DTNet \cite{mao2024dielectric} incorporates a pretrained module (PFP) to enhance predictive accuracy under limited dielectric data conditions. Inspired by this strategy, we conducted experiments to assess the impact of pretraining on GoeCTP. As illustrated in Figure \ref{fig:overview}, GoeCTP was first pretrained on a large-scale dataset\footnote{{
It is publicly available as \texttt{megnet2} within the Python package (jarvis-tools) \cite{choudhary2020joint}.
}} from the Materials Project, which includes 133,000 crystal structures with their corresponding formation energies. Following pretraining, the scalar prediction readout head was replaced with a tensor prediction head, and the network was subsequently fine-tuned on approximately 5,000 dielectric tensor samples.
The comparative results between training from scratch and fine-tuning are shown in Figure \ref{dielectric_MP_finetuning}. The results reveal that all three GoeCTP variants achieve lower prediction errors after fine-tuning. Specifically, the average MAE of GoeCTP decreased from 1.411 to 1.381, that of GoeCTP (Mat.) decreased from 1.441 to 1.359, and that of GoeCTP (iCom.) decreased from 1.414 to 1.350. Notably, after fine-tuning, both GoeCTP (Mat.) and GoeCTP (iCom.) outperform the base GoeCTP. These findings indicate that more complex architectures benefit more substantially from pretraining, as they are otherwise more prone to overfitting when trained on limited datasets.

\begin{figure*}[t] 
  \centering   
  \includegraphics[width=\textwidth]{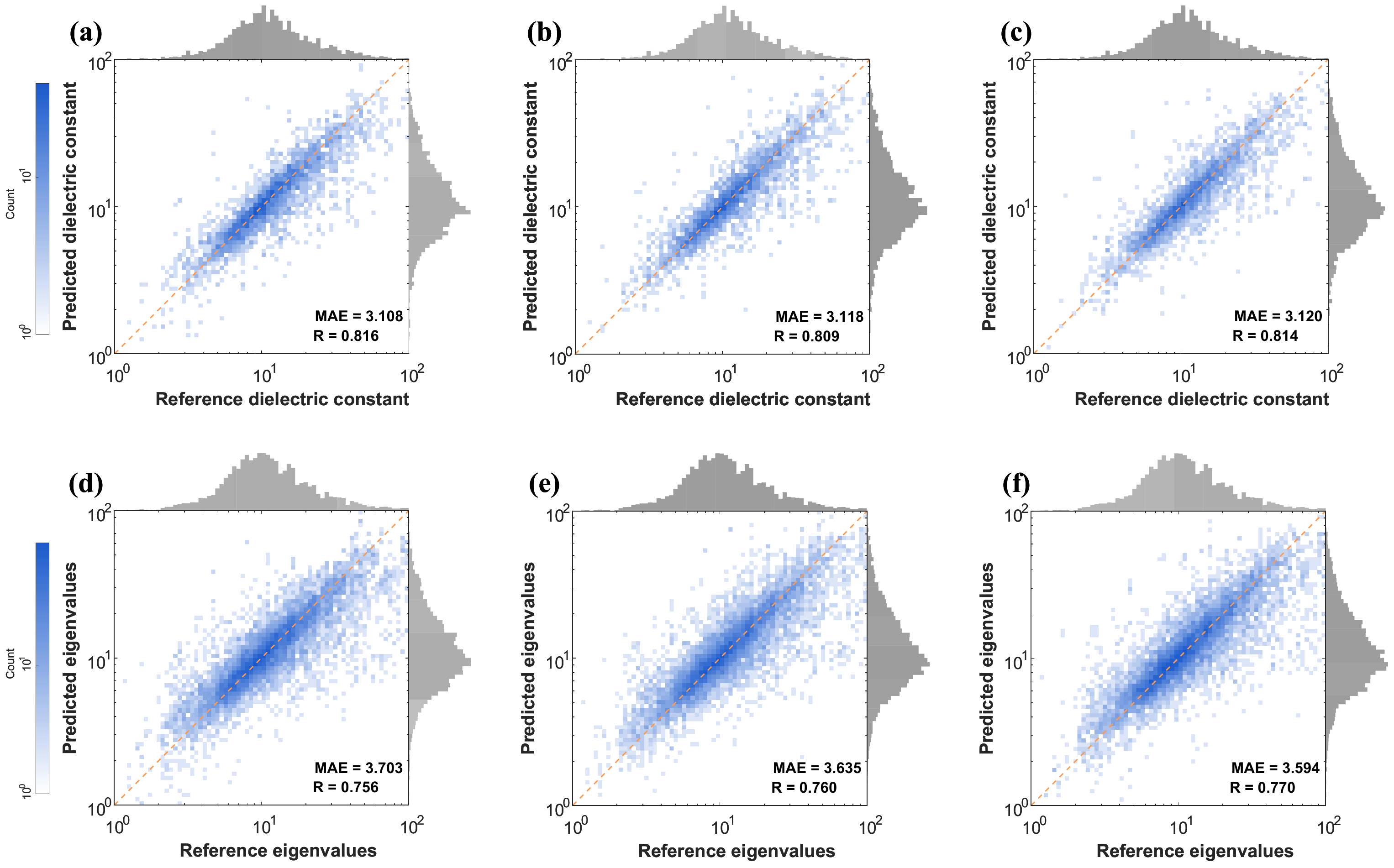}
  \caption{Performance of GoeCTP models on the test split of the Materials Project dataset for two prediction tasks. Panels (a), (b), and (c) show the results of GoeCTP, GoeCTP (Mat.), and GoeCTP (iCom.) respectively on the polycrystalline dielectric constant prediction task. Panels (d), (e), and (f) present the corresponding results of GoeCTP, GoeCTP (Mat.), and GoeCTP (iCom.) on the tensor eigenvalue prediction task.
  } 
  \label{eigenvalues}
\end{figure*}

Following \citet{mao2024dielectric}, we further evaluated the performance of GoeCTP by computing polycrystalline dielectric constants and tensor eigenvalues from the predicted dielectric tensors, as these quantities are crucial for identifying high-dielectric and highly anisotropic materials. On the Materials Project dataset, we repeated each experiment five times using different random data splits. The predicted tensors from all test sets were aggregated, and the corresponding polycrystalline dielectric constants and tensor eigenvalues were subsequently computed. As illustrated in Figure \ref{eigenvalues}, the predictions generated by the GoeCTP family exhibit strong consistency with the DFT reference values. GoeCTP, GoeCTP (Mat.), and GoeCTP (iCom.) perform comparably in predicting polycrystalline dielectric constants, while GoeCTP (iCom.) shows clear superiority in predicting tensor eigenvalues.

\begin{figure*}[t] 
  \centering   
  \includegraphics[width=\textwidth]{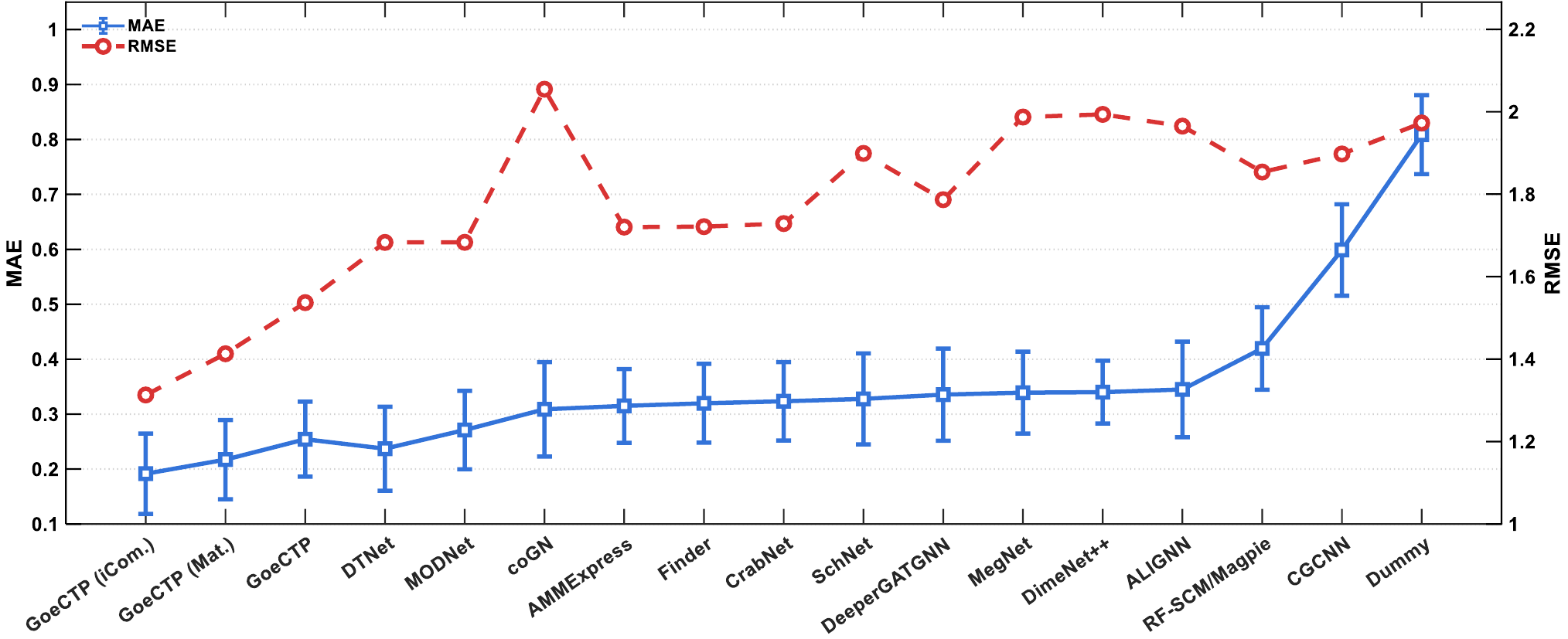}
  \caption{Benchmarking our approach against other algorithms on the Matbench dataset. 
  } 
  \label{Matbench}
\end{figure*}

To further validate the accuracy of our approach, we employed the Matbench dataset, which is a curated subset of the Materials Project structures, to evaluate model performance on the dielectric tensor prediction task. In this setting, we assessed the refractive indices derived from the dielectric tensors predicted by GoeCTP, following the evaluation protocol described in \citet{mao2024dielectric}. Figure \ref{Matbench} presents the comparison of MAE values together with their standard deviations, as well as the average root mean square error (RMSE) across five independent runs under the official Matbench evaluation pipeline.
For the dielectric prediction task, GoeCTP achieves performance comparable to DTNet, with only a slightly higher MAE. In contrast, both GoeCTP (Mat.) and GoeCTP (iCom.) outperform DTNet and all other methods currently listed on the official Matbench leaderboard, thereby establishing new state-of-the-art results for dielectric tensor prediction.

\subsection{Discovery of dielectric materials}
\label{Discovery of dielectric materials}

\textbf{Preparation of candidate structures. }
The previous sections have demonstrated that GoeCTP is highly effective in predicting dielectric tensors. In this section, we apply GoeCTP to perform virtual screening. To construct the candidate set, we retrieved 13,004 materials from the Materials Project database that satisfy the criteria of having nonzero band gaps and $E_{\text{hull}} = 0$, where $E_{\text{hull}}$ denotes the energy above the convex hull. These constraints ensure thermodynamic stability while excluding metallic compounds, thereby restricting the set to stable materials. There is no overlap between this candidate set and our training dataset. 
Our objective is to utilize GoeCTP to conduct large-scale virtual screening across this candidate set in order to identify materials that exhibit either high dielectric performance or strong anisotropy. The detailed procedure for screening is described in the Methods section.

\textbf{High-dielectric materials.} 
Dielectric materials are indispensable functional components in modern information, communication, and energy technologies~\citep{hao2013review}. For instance, materials with a high dielectric constant (high-$\kappa$) are particularly valuable because they enable higher energy-density capacitors and allow further device miniaturization without increasing power consumption.
At the same time, a wide electronic band gap $E_g$ is crucial for maintaining low leakage currents under strong electric fields~\citep{kittl2009high,yim2015novel,mao2024dielectric}. 
Therefore, novel dielectric materials should simultaneously exhibit a high dielectric constant and a large band gap, which also serve as the key criteria in our screening process.

To represent the capacity of electric energy storage, the dielectric tensor $\boldsymbol{\varepsilon}$ can be simplified into a scalar quantity~\citep{mao2024dielectric} as follows:
\begin{equation}
\label{dielectric_constant}
\overline{{\varepsilon}}= \frac{1}{3}\sum_{i=1}^{3} \lambda_i,
\end{equation}
where $\lambda_i$ is the eigenvalue of $\boldsymbol{\varepsilon}$ and $\overline{\varepsilon}$ is the dielectric constant.
We performed virtual screening by employing GoeCTP to predict the dielectric tensor $\boldsymbol{\varepsilon}$ for all structures in the candidate set and subsequently computed the scalar dielectric constant $\overline{\varepsilon}$ from these predictions. To reduce the computational cost of post-screening DFT validation, we restricted the virtual screening to GoeCTP (iCom.) and GoeCTP, which demonstrated superior performance according to the results in Table \ref{dielectric_MP}. 

\begin{figure*}[t] 
  \centering   
  \includegraphics[width=\textwidth]{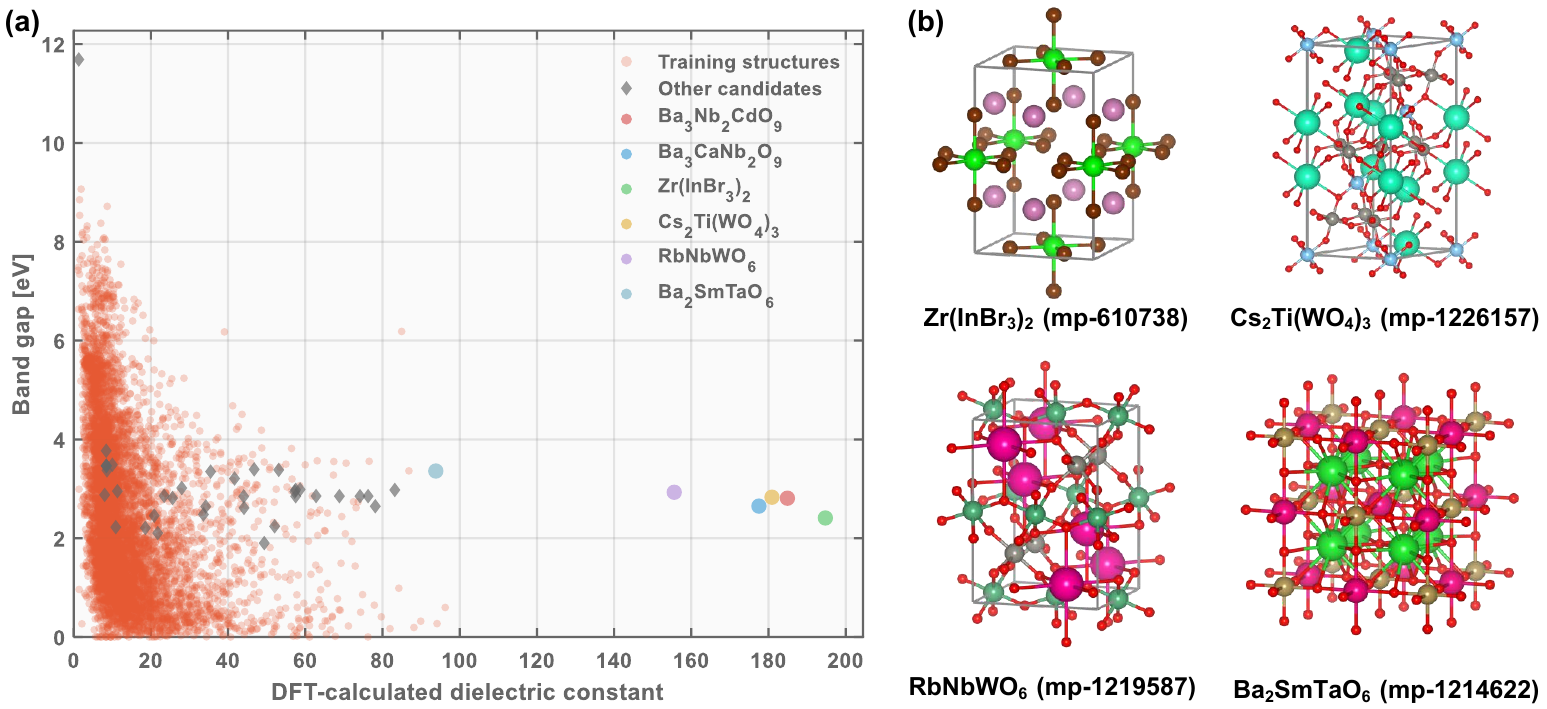}
  \caption{\textbf{ML-predicted and DFT-calculated dielectric constants for the candidate
 materials}. (a) DFT-calculated  $\overline{\varepsilon}$-$E_g$ data points for all stable structures in MP dataset and new stable
 structures screened by GoeCTP. (b) Structure visualization of the materials with a high dielectric constant as validated by DFPT calculations.
  } 
  \label{screening}
\end{figure*}

Figure \ref{screening} presents the distribution of the band gap $E_g$ and $\overline{\varepsilon}$ for materials in both the training set and the screened candidates. All materials in the MP dataset exhibit dielectric constants below 100. In contrast, GoeCTP trained from scratch and GoeCTP after fine-tuning both identified a promising new material, Ba$_2$SmTaO$_6$ (mp-1214622, $E_g = 3.36$ eV, $\varepsilon = 93.81$), which achieves a superior combination of band gap and dielectric constant compared with existing materials in the MP dataset. This compound was also reported by DTNet through virtual screening.
GoeCTP (iCom.) achieved even stronger screening performance. When trained from scratch, GoeCTP (iCom.) identified three potential new materials: Ba$_2$SmTaO$_6$ (mp-1214622, $E_g = 3.36$ eV, $\overline{\varepsilon} = 93.81$), Cs$_2$Ti(WO$_4$)$_3$ (mp-1226157, $E_g = 2.83$ eV, $\overline{\varepsilon} = 180.89$), and RbNbWO$_6$ (mp-1219587, $E_g = 2.93$ eV, $\overline{\varepsilon} = 155.57$). These three materials were also identified by DTNet, but only through active-learning-based screening. In contrast, GoeCTP (iCom.) discovered them directly without active learning, demonstrating its clear advantage over DTNet.
Moreover, GoeCTP (iCom.) after fine-tuning not only rediscovered these materials but also revealed an additional new candidate, Zr(InBr$_3$)$_2$ (mp-610738, $E_g = 2.41$ eV, $\overline{\varepsilon} = 194.72$), which possesses an even higher dielectric constant. This finding highlights the benefit of pretraining in improving the model’s capacity to discover novel high-performance dielectrics.
Additionally, GoeCTP and GoeCTP (iCom.) identified two candidate materials that have already been reported. GoeCTP discovered Ba$_3$Nb$_2$CdO$_9$ (mp-1214502, $E_g = 2.81$ eV, $\overline{\varepsilon} = 184.89$), while GoeCTP (iCom.) identified both Ba$_3$Nb$_2$CdO$_9$ and Ba$_3$CaNb$_2$O$_9$ (mp-1214569, $E_g = 2.65$ eV, $\overline{\varepsilon} = 177.51$). These materials were not included in the MP dataset used in Section \ref{Materials_Project_Dataset} but are reported on the official MP website. Their identification further demonstrates the capability of GoeCTP to discover high-dielectric materials.
Because the pretrained PFP module in DTNet imposes restrictions on the allowable element types, certain high-dielectric candidates such as Ba$_3$Nb$_2$CdO$_9$ and Ba$_3$CaNb$_2$O$_9$ cannot be recognized by DTNet. In contrast, GoeCTP is free from such elemental constraints.

\textbf{Highly anisotropic materials.}
Anisotropic materials can provide directional responses to external stimuli that isotropic materials cannot~\citep{lou2025discovery}. Materials exhibiting large anisotropy ratios are considered promising candidates for polarization-sensitive devices, optical fiber sensors, and energy-related applications~\citep{wang2022anisotropic,tudi2022potential,oberg2014control,li2021demonstration}.
In this task, we use the anisotropy ratio $\alpha_r = \lambda_{\text{max}} / \lambda_{\text{min}}$, defined as the ratio between the maximum and minimum eigenvalues of the dielectric tensor~\citep{lou2025discovery}, as the quantitative metric for evaluating dielectric anisotropy. 

After performing virtual screening using GoeCTP (iCom.) and GoeCTP, the GoeCTP (iCom.) model trained from scratch identified a potential high-anisotropy material, SeI$_2$ (mp-861871, $\alpha_r = 96.763$), which was also reported by DTNet.
In contrast, the fine-tuned GoeCTP (iCom.) only discovered ClO$_3$ (mp-22869, $\alpha_r = 22.8204$), the GoeCTP model trained from scratch identified HfSe$_2$ (mp-985831, $\alpha_r = 6.154$), and the fine-tuned GoeCTP identified Y$_2$Cl$_2$ (mp-1206803, $\alpha_r = 6.371$).
However, these materials are either molecular crystals or exhibit low anisotropy ratios, making them less suitable for practical applications~\citep{mao2024dielectric}.
For the task of identifying highly anisotropic materials, the comparison between GoeCTP (iCom.) and GoeCTP indicates that more complex network architectures may be advantageous for discovering materials with strong anisotropy.
Additionally, the results from GoeCTP (iCom.) suggest that the pretraining and fine-tuning strategy may reduce the network’s ability to identify highly anisotropic candidates in practice.
Compared with DTNet, GoeCTP (iCom.) successfully discovered the potential highly anisotropic material SeI$_2$ without relying on active learning, demonstrating its superior capability in anisotropic material discovery.

\section{Discussion}
Our proposed GoeCTP framework establishes a general and unified approach for predicting material dielectric tensor properties. Conventionally, EGNNs preserve equivariance by employing carefully designed message-passing mechanisms, such as encoding interatomic edge vectors or other equivariant geometric features. During message passing, to maintain the equivariant nature of these representations, operations are typically restricted to linear transformations or gated equivariant nonlinearities. These architectural constraints often limit the expressive capacity of the network and may consequently reduce predictive accuracy.
In contrast, GoeCTP achieves equivariance through frame averaging, which imposes no constraints on message-passing design. This flexibility enables arbitrary customization of the message-passing module and allows the incorporation of nonlinear transformations to enhance the model’s representational and fitting capacity, thereby improving predictive performance. As demonstrated in Section \ref{Materials_Project_Dataset}, our method consistently achieves stable and reliable performance improvements on both the Matbench and Materials Project datasets.

Moreover, frame averaging facilitates a more flexible pretraining strategy. As discussed in Section \ref{Materials_Project_Dataset}, when pretraining on large scalar datasets such as those containing formation energies, the network must maintain invariance rather than equivariance. The DTNet adopts the PFP module, which effectively pretrains only part of the non-equivariant layers before fine-tuning. Directly transferring such pretrained models to tensor prediction tasks can reduce generalization capability due to the mismatch in symmetry constraints. By integrating frame averaging, we can pretrain the entire backbone network and later embed it into an equivariant framework during fine-tuning, thereby achieving physically consistent tensor predictions.
Additionally, DTNet employs the PFP module, which is a proprietary commercial component. This restricts the range of elemental compositions that can be processed, potentially excluding some high-performance materials. For example, as shown in Section \ref{Discovery of dielectric materials}, the potential high-dielectric materials Ba$_3$Nb$_2$CdO$_9$ (mp-1214502, $E_g = 2.81$ eV, $\overline{\varepsilon} = 184.89$) and Ba$_3$CaNb$_2$O$_9$ (mp-1214569, $E_g = 2.65$ eV, $\overline{\varepsilon} = 177.51$) could not be recognized by DTNet due to its elemental limitations. In contrast, GoeCTP is easy to implement and capable of unrestricted pretraining, demonstrating clear advantages in flexibility and accessibility.

Although the accuracy of GoeCTP remains constrained by the limited size of available datasets, as illustrated in Figure \ref{eigenvalues}, we have already demonstrated in Section \ref{Discovery of dielectric materials} that GoeCTP can effectively identify potential novel materials. This finding underscores the promise of GoeCTP frameworks for materials discovery. Future research directions include further enhancing model accuracy under data-scarce conditions by leveraging GoeCTP’s flexibility and scalability, such as incorporating more advanced backbone architectures or integrating large-scale pretrained models to improve dielectric tensor prediction performance.

In conclusion, this work presents GoeCTP, a scalable and flexible framework for dielectric tensor prediction that decouples O(3) equivariance from backbone architecture design through frame averaging. By removing the need for specially constrained equivariant message-passing operations, GoeCTP enables the use of expressive non-equivariant backbones while rigorously preserving the symmetry requirements of tensorial properties. Extensive experiments on JARVIS-DFT, Materials Project, and Matbench datasets demonstrate that GoeCTP achieves competitive or superior performance compared to state-of-the-art equivariant models, while offering improved flexibility in pretraining, fine-tuning, and large-scale screening. Moreover, its successful application to virtual screening highlights its practical value in identifying novel high-dielectric and highly anisotropic materials.

\section{Methods}

\subsection{Equivariance}
Atomic systems are typically represented by the types of atoms and their coordinates in three-dimensional Euclidean space.
Under the influence of symmetries such as rotations, translations, and reflections, the configuration of the system may undergo transformations. Scalar properties of atomic systems, such as total energy, remain invariant under O(3) group transformations in 3D space. This invariance implies that these properties retain the same values regardless of the system's orientation.
However, tensor properties such as atomic forces, stress tensors, and dielectric tensors vary consistently with changes in the system's orientation. When the atomic configuration is rotated or reflected, these tensor transform accordingly to preserve their physical relationship with the new spatial orientation of the system.

In the context of dielectric tensor prediction, it is desirable for the model to produce outputs that comply with the physical constraints imposed by such O(3) group transformations. This ensures that the model generalizes effectively to arbitrary orientations of the same underlying crystal structure. Formally, let the atomic positions of a crystal structure be denoted by 
$\mathbf{X}$, and let $f_\theta(\cdot)$ represent the prediction model with $f_\theta(\mathbf{X})=\boldsymbol{\varepsilon}$. When the input structure undergoes an orthogonal transformation, the model output is expected to transform in a physically consistent manner as follows:
\begin{equation}
\label{equivariance_eq}
f_\theta(\mathbf{R}\mathbf{X}) = \mathbf{R}f_\theta(\mathbf{X})\mathbf{R}^\top,
\end{equation}
where $\mathbf{R}\in\mathbb{R}^{3 \times 3}$ is an arbitrary O(3) group transformation. 

\subsection{Model Architectures}

\textbf{Overview.}
Our proposed model consists of three primary components: the frame-averaging module, the message-passing module, and the readout module. A crystal material is represented as a graph $G=(V, E)$. We adopt the multi-edge graph construction for crystals from CGCNN \cite{xie2018crystal} to obtain $G=(V,E)$.
The node set is defined as $V=\{v^{(l)}_i|v^{(l)}_i \in\mathbb{R}^{d_a \times N} \}$, where each node corresponds to an atom described by a $d_a$-dimensional feature vector. Here, $v^{(l)}_i$ denotes the atomic feature vector of the $i$-th atom in the $l$-th layer.
The edge set is defined as $E=\{e^{(h)}_{ij}|e_{ij} \in\mathbb{R}^{d_e \times H} \}$, where each edge feature $e^{(h)}_{ij}$ represents a $d_e$-dimensional vector corresponding to the $h$-th edge between atoms $i$ and $j$. The edge features are typically constructed based on the Euclidean distances between atomic pairs.
In the subsequent sections, we detail the design and functionality of each of these three components.


%
\textbf{Frame averaging.}
Frame averaging has been widely adopted to enforce equivariance and invariance in geometric deep learning \citep{linequivariance, ma2024canonicalization}. In the context of dielectric tensor prediction, a frame can be interpreted as an orthogonal matrix $\mathbf{F}\in\mathbb{R}^{3 \times 3}$, deriving from an O(3)-equivariant map denoted as $\mathbf{F}=h(\mathbf{X})$.
For an arbitrary non-equivariant prediction model $g_\theta(\cdot)$, the frame averaging approach can be used to construct an equivariant prediction model $f_\theta(\cdot)$ that satisfies the equivariance condition in Equation \ref{equivariance_eq}, as formalized in Equation \ref{frame_qian}.
This formulation decouples the requirement of equivariance from the architectural design of the neural network, allowing message-passing schemes to be flexibly designed without additional symmetry constraints.
In this work,
we employ polar decomposition \citep{higham1986computing,jiaospace,hall2013lie} as $h(\cdot)$ to compute the frame $\mathbf{F}$ in
Equation \ref{frame_qian}. Polar decomposition is inherently O(3)-equivariant and, as an algebraic method, introduces negligible computational overhead.
We use the lattice matrix $\mathbf{L} \in \mathbb{R}^{3\times3}$ as the input, since it is an invertible matrix that can be uniquely decomposed via polar decomposition as $\mathbf{L} = \mathbf{F}\mathbf{H}$, where $\mathbf{F} \in \mathbb{R}^{3\times3}$ is the orthogonal matrix serving as the frame and $\mathbf{H} \in \mathbb{R}^{3\times3}$ is a positive semi-definite matrix.

\textbf{Message passing module.}
The message-passing module, together with the readout module, constitutes $g_\theta(\cdot)$ in Equation~\ref{frame_qian}.
Different message-passing strategies are adopted for GoeCTP, GoeCTP (Mat.), and GoeCTP (iCom.) to investigate the effect of architectural complexity on model performance.
For the base GoeCTP model, the message passing is formulated as follows:
\begin{equation}
\begin{aligned}
& \boldsymbol{msg}_{ij}=\sum_{h}\boldsymbol{\xi}(\operatorname{LN} (v^{(l)}_i)|\operatorname{LN} (v^{(l)}_j)|\operatorname{LN}_e (e^{(h)}_{ij})), \\
&v^{(l+1)}_i=\operatorname{softplus}(v^{(l)}_i+\mathrm{BN}(\sum_{j\in\mathcal{N}_i}\boldsymbol{msg}_{ij})), \\
\end{aligned}
\end{equation}
where $\mathrm{LN}(\cdot)$ and $\mathrm{LN}_e(\cdot)$ denote the linear transformation layers, $\boldsymbol{\xi}(\cdot)$ represents a nonlinear transformation function, $\operatorname{BN}(\cdot)$ refers to the batch normalization, and $|$ denote the concatenation. 

For GoeCTP (Mat.), we employ the transformer-based message passing scheme proposed in Matformer \cite{yan2022periodic}.
Specifically, for each node pair $(i, j)$, the message from node $j$ to node $i$ is first transformed into query, key, and value representations defined as $\boldsymbol{q}_{ij}=\mathrm{LN}_q(v^{(l)}_i)$, $\boldsymbol{k}_{ij}^{(h)}=(\mathrm{LN}_k(v^{(l)}_i)|\mathrm{LN}_k(v^{(l)}_j)|\mathrm{LN}_e(e^{(h)}_{ij}))$, and $\boldsymbol{v}_{ij}^{(h)}=(\mathrm{LN}_v(v^{(l)}_i)|\mathrm{LN}_v(v^{(l)}_j)|\mathrm{LN}_e(e^{(h)}_{ij}))$, where $\mathrm{LN}_q(\cdot)$, $\mathrm{LN}_k(\cdot)$, $\mathrm{LN}_v(\cdot)$, $\mathrm{LN}_e(\cdot)$ denote the linear transformations. The self-attention mechanism is then computed as:
\begin{equation}
\begin{aligned}
\boldsymbol{\alpha}^{(h)}_{ij}= \frac{\boldsymbol{q}_{ij}\circ\boldsymbol{\xi}_k(\boldsymbol{k}^{(h)}_{ij})}{\sqrt{d_{\boldsymbol{q}_{ij}}}}&,
\boldsymbol{msg}_{ij}=\sum_{h}\operatorname{sigmoid}(\operatorname{BN}(\boldsymbol{\alpha}_{ij}^{(h)}))\circ\boldsymbol{\xi}_v(\boldsymbol{\upsilon}_{ij}^{(h)}), \\
v^{(l+1)}_i=&\operatorname{softplus}(v^{(l)}_i+\mathrm{BN}(\sum_{j\in\mathcal{N}_i}\boldsymbol{msg}_{ij})),\\
\end{aligned}
\end{equation}
where $\boldsymbol{\xi}_k$ and $\boldsymbol{\xi}_v$ denote nonlinear transformations applied
to key and value features, respectively. The operators $\circ$ indicates
the Hadamard product, and $\sqrt{d_{\boldsymbol{q}_{ij}}}$ represents the dimensionality of $\boldsymbol{q}_{ij}$. 

For GoeCTP (iCom.), we extend GoeCTP (Mat.) by introducing an additional edge-wise transformer layer to dynamically update the edge features, following the design of iComFormer \cite{yancomplete}. For each edge feature $e^{(h)}_{ij}$, we first use the lattice matrix $\mathbf{L}=[\boldsymbol{l}_1,\boldsymbol{l}_2,\boldsymbol{l}_3]$ to compute the high-dimensional angle features $\theta^{(h)}_{ijm}$ corresponding to the three angles formed between the edge and the three lattice vectors, with $m=1,2,3$. Simultaneously, each lattice vector $\boldsymbol{l}_m$ is transformed into high-dimensional features $v^{{l}}_m$, and all features are constructed using radial basis functions~\cite{schutt2017schnet}.
We then compute the query, key, and value representations as $\boldsymbol{q}^{(h)}_{ij}=\mathrm{LN}_q(e^{(h)}_{ij})$, $\boldsymbol{k}_{ijm}^{(h)}=(\mathrm{LN}_k(e^{(h)}_{ij})\mid \mathrm{LN}^m_k(v^{{l}}_m))$, 
and $\boldsymbol{v}_{ijm}^{(h)}=(\mathrm{LN}_v(e^{(h)}_{ij})\mid \mathrm{LN}^m_v(v^{{l}}_m))$, where $\mathrm{LN}_q(\cdot)$, $\mathrm{LN}_k(\cdot)$, $\mathrm{LN}_v(\cdot)$, $\mathrm{LN}^m_k(\cdot)$, and $\mathrm{LN}^m_v(\cdot)$ denote linear transformations. 
The updated edge feature $e^{(h)*}_{ij}$ is obtained through:
\begin{equation}
\begin{aligned}
\boldsymbol{\alpha}^{(h)}_{ijm}&= \frac{\boldsymbol{q}^{(h)}_{ij}\circ\boldsymbol{\xi}_k(\boldsymbol{k}^{(h)}_{ijm}|\mathrm{LN}_\theta(\theta^{(h)}_{ijm}))}{\sqrt{d_{\boldsymbol{q}^{(h)}_{ij}}}},\\
\boldsymbol{msg}_{ij}=\sum_{m}&\operatorname{sigmoid}(\operatorname{BN}(\boldsymbol{\alpha}_{ijm}^{(h)}))\circ\boldsymbol{\xi}_v((\boldsymbol{v}^{(h)}_{ijm}|\mathrm{LN}_\theta(\theta^{(h)}_{ijm}))), \\
e^{(h)*}_{ij}&=\operatorname{softplus}(e^{(h)}_{ij}+\mathrm{BN}(\boldsymbol{msg}_{ij})),\\
\end{aligned}
\end{equation}
where $\boldsymbol{\xi}_k$ and $\boldsymbol{\xi}_v$ represent nonlinear transformations, and $\mathrm{LN}_\theta(\cdot)$ represents a linear mapping applied to the angular features.

\textbf{Readout.}
The final node representations $v_i^\mathrm{final}$ are aggregated to predict the tensorial property as follows:
\begin{equation}
\boldsymbol{\varepsilon}=\operatorname{MLP}(\sum_{i}v_i^\mathrm{final}),
\end{equation}
where $\operatorname{MLP}(\cdot)$ represents a multilayer perceptron that maps the aggregated representation to a tensor output of the desired dimensionality.
Subsequently, the frame $\mathbf{F}$ obtained from the polar decomposition is used to transform the predicted tensor $\boldsymbol{\varepsilon}$ into its final equivariant form:
\begin{equation}
\boldsymbol{\varepsilon}^\mathrm{final}=\mathbf{F}\boldsymbol{\varepsilon}\mathbf{F}^\top. 
\end{equation}

This transformation ensures that the model output strictly satisfies the O(3)-equivariance condition.

\subsection{Hyperparameters}

For the JARVIS-DFT dataset and the Materials Project dataset, model training was conducted using the Huber loss and the AdamW optimizer \citep{loshchilovdecoupled}, with a weight decay of $10^{-5}$ and a polynomial learning rate decay schedule. The initial learning rate was set to 0.001, with training performed for 200 epochs and a batch size of 64.
In the pretraining–fine-tuning experiments on the Materials Project dataset, the pretraining stage was configured with a learning rate of 0.001 for 500 epochs, while the fine-tuning stage employed a learning rate of 0.01 for 40 epochs. All other hyperparameter settings remained consistent with those used in the Materials Project experiments.
For network configuration, GoeCTP employed a single message passing layer, whereas GoeCTP (Mat.) and GoeCTP (iCom.) utilized four message passing layers. Additionally, GoeCTP (iCom.) incorporated an edge-wise transformer layer immediately after the first message passing layer.
During crystal graph construction, the cutoff radius was determined by the 16th nearest neighboring atom. Across all GoeCTP variants, each message passing layer transformed 128-dimensional input features into 128-dimensional output features.

\subsection{Virtual screening}



We performed virtual screening using five independently trained models, each corresponding to a different random data split. To ensure a fair comparison with the screening results reported for DTNet \cite{mao2024dielectric}, candidate selection was based on a weighted combination of the mean predictions and standard deviations from these five models. This strategy resembles the active learning approach adopted in DTNet but uses only the results from the first iteration, without multi-round refinement.
For each screening task, the top 60 candidates with the highest polycrystalline dielectric constants and anisotropy ratios were selected, respectively. Considering computational budget limitations, only candidates with unit cell sizes of up to 20 atoms were validated using density functional perturbation theory (DFPT). In the polycrystalline dielectric constants task, validation was further restricted to candidates exhibiting band gaps greater than 1.8 eV.

\subsection{DFPT configuration}
First-principles calculations of dielectric tensors of novel materials discovered in this work were conducted by Vienna ab initio simulation package (VASP) \cite{kresse1996efficiency}, in which a projector-augmented wave potential \cite{kresse1999ultrasoft} and a 700 eV cut-off energy was employed.
Perdew–Burke–Ernzerhof (PBE) functional \cite{perdew1996generalized} was adopted to describe exchange and correlation interactions in the framwork of generalized gradient approximation.
For materials containing transition metal elements, Hubbard U corrections were considered following the general settings in high-throughput calculations in Materials Project (MP) \cite{jain2013commentary,petousis2017high}.
The k-point density was set with a Gamma-centered Monkhorst–Pack grid \cite{monkhorst1976special} with K-spacing equal to 0.03$\times$2$\pi$ to enable a refined sampling of the Brillouin-zone for both electronic and ionic contribution calculation.
A Gaussian smearing scheme with a width of 0.05 eV was employed to mitigate the discontinuity in orbital occupations near the Fermi level.
In all calculations, following settings in previous works \cite{petousis2017high,zhang2024local,lou2025discovery}, the input crystal structures were not optimized and directly used for computing dielectric properties.
During self-consistency calculations, the energy difference threshold was set as 1$\times$10$^{-8}$ eV for both electronic and ionic contribution calculation, and a -0.001 eV/\AA\ force threshold was additionally included in ionic contribution calculation.

\bibliographystyle{unsrtnat}
\bibliography{main}

\end{document}